\documentclass[%
superscriptaddress,
 amsmath,amssymb,
 aps,
prb,
twocolumn
]{revtex4}

\usepackage{graphicx}
\usepackage{dcolumn}
\usepackage{bm}
\usepackage{float}
\usepackage{mathrsfs}
\usepackage{amsmath}
\usepackage{color}
\usepackage{ulem}

\begin{document}


\title{Nearly hyperuniform, nonhyperuniform, and antihyperuniform density fluctuations in two-dimensional transition metal dichalcogenides with defects}

\author{Duyu Chen}
\email[correspondence sent to: ]{duyu@alumni.princeton.edu}
\affiliation{Tepper School of Business, Carnegie Mellon
University, Pittsburgh, PA 15213, United States}
\altaffiliation[Present address: ]
{Materials Research Laboratory, University of California, Santa Barbara, California 93106, United States}

\author{Yu Zheng}
\affiliation{Department of Physics, Arizona State University,
Tempe, AZ 85287, United States}

\author{Chia-Hao Lee}
\affiliation{Department of Materials Science and Engineering, University of Illinois Urbana-Champaign, Urbana, IL 61801, United States}

\author{Sangmin Kang}
\affiliation{Semiconductor Research Center, Samsung Electronics, Hwaseong-si, Gyeonggi-do 445701, South Korea}

\author{Wenjuan Zhu}
\affiliation{Department of Electrical and Computer Engineering, University of Illinois Urbana-Champaign, Urbana, IL 61801, United States}

\author{Houlong Zhuang}
\affiliation{Mechanical and Aerospace Engineering, Arizona State
University, Tempe, AZ 85287, United States}

\author{Pinshane Y. Huang}
\affiliation{Department of Materials Science and Engineering, University of Illinois Urbana-Champaign, Urbana, IL 61801, United States}

\author{Yang Jiao}
\affiliation{Materials Science and Engineering, Arizona State
University, Tempe, AZ 85287, United States}
\affiliation{Department of Physics, Arizona State University,
Tempe, AZ 85287, United States}
\date{\today}

\begin{abstract}
Hyperuniform many-body systems in $d$-dimensional Euclidean space $\mathrm{R}^d$ are characterized by completely suppressed
(normalized) infinite-wavelength density fluctuations, and appear to be endowed with novel exotic physical properties. Recently, 
hyperuniform systems of disordered varieties have been observed in the context of various atomic-scale two-dimensional (2D) materials. In this work, we analyze the effects of localized defects on the density fluctuations across length scales and on the hyperuniformity property of \textit{experimental} samples of two-dimensional transition metal dichalcogenides. In particular, we extract atomic coordinates from time series annular dark field-scanning transmission electron microscopy (ADF-STEM) imaging data of 2D tungsten chalcogenides with the 2H structure (Te-doped 2H-WSe$_2$) showing continuous development and evolution of electron-beam induced defects, and construct the corresponding chemical-bonding informed coordination networks between the atoms. We then compute a variety of pair statistics and bond-orientational statistics to characterize the samples. At low defect concentrations, the corresponding materials are nearly hyperuniform, characterized by significantly suppressed scattering at the zero wave-number limit (omitting forward/ballistic scattering). As more defects are introduced, the (approximate) hyperuniformity of the materials is gradually destroyed, and the system becomes non-hyperuniform even when the material still contains a significant amount of crystalline regions. At high defect concentrations, the structures become antihyperuniform with diverging (normalized) large-scale density fluctuations, mimicking those typically observed at the thermal critical points associated with phase transitions. Overall, the defected materials possess varying degrees of orientation order, and there is apparently no intermediate hexatic phase emerging. To understand the observed nearly hyperuniform density fluctuations in the slightly defected materials, we construct a minimalist structural model and demonstrate that the experimental samples can be essentially viewed as perturbed honeycomb crystals with small correlated displacements and double chalcogen vacancies. Moreover, the small correlated displacements alone can significantly degrade hyperuniformity of the perfect honeycomb structure. Therefore, even a small amount of vacancies, when coupled with correlated displacements, can completely destroy hyperuniformity of the system. 
\end{abstract}


\maketitle


\section{Introduction}

Hyperuniformity is a recently introduced novel concept that provides a unified framework to categorize
crystals, quasicrystals, and certain unusual disordered systems \cite{To03, To16b, To18a}. Hyperuniform many-body systems in $d$-dimensional Euclidean space $\mathrm{R}^d$ are characterized by completely suppressed
(normalized) density fluctuations at large length scales. In particular, the
static structure factor $S(k)$, which is proportional to 
the scattering intensity in $X$-ray, light, or neutron scattering experiments, vanishes in the infinite-wavelength (or
zero-wavenumber) limit for hyperuniform systems, i.e., $\lim_{k\rightarrow 0}S(k) = 0$,
where $k$ is the wavenumber. Here $S(k)$ is defined as
\begin{equation}
\label{eq_1}
S(k) \equiv 1 + \rho\Tilde{h}(k), 
\end{equation}
where $\rho$ is the number density of the system, and $\Tilde{h}(k)$ is the Fourier transform of the total correlation
function 
\begin{equation}
\label{eq_2}
h(r) = g_2(r) - 1, 
\end{equation}
where $g_2(r)$ is the pair correlation function. 
Note that this definition implies that the forward scattering contribution to the
diffraction pattern is omitted. Equivalently, the local number variance 
\begin{equation}
\label{eq_3}
\sigma_N^2(R)\equiv \langle N^2(R)\rangle - \langle N(R) \rangle^2
\end{equation}
of these systems associated
with a spherical window of radius $R$ grows more slowly than
the window volume (i.e., a scaling of $R^d$ in $d$-dimensional
Euclidean space) in the large-$R$ limit \cite{To03, To18a}, where $N(R)$ is the number of particles in a spherical window with radius $R$ randomly placed into the system. Recently, hyperuniformity has been observed in many physical, biological and material systems \cite{Do05, Za11a, Za11b, Ba08, Ba09, Ji11, Le83, Dr15, He15, He13, Ga02, He17a, He17b, We17, Di18b, Le19a, Le19b, Ru19, San19, San20}.

Hyperuniform systems, in particular the disordered varieties, appear to be endowed with novel photonic, phononic, transport and mechanical properties, and wave-propagation characteristics \cite{Fl09, Ma13, Gk17, Zh16, Xu17, Ch18a, To18b}. For example, disordered hyperuniform dielectric networks were found to possess complete photonic band gaps comparable in size to photonic crystals, while at the same time maintaining statistical isotropy, enabling waveguide geometries not possible with photonic crystals \cite{Fl09, Ma13}. Moreover, disordered hyperuniform patterns can have nearly optimal color-sensing capabilities, as evidenced by avian photoreceptors \cite{Ji14}. In addition, disordered stealthy hyperuniform two-phase materials and cellular solids were recently found to possess virtually optimal transport properties \cite{Zh16, Ch18a, To18b}. The reader is referred to Ref. \cite{To18a} for a thorough overview of hyperuniform systems.

Despite the exotic structural characteristics and physical properties that hyperuniform systems possess, in practice it is very difficult to find perfect hyperuniform systems of both ordered and disordered varieties due to the inevitable existence of imperfection \cite{Ki18}, except in a few cases where tailored optimization techniques are designed and deployed to fabricate perfect hyperuniform materials \cite{To15, Zh15a, Di18a, Ki19a, Ki19b}. Therefore, it is important to systematically investigate how various types of imperfections affect hyperuniformity and the associated physical properties of the systems. In the past, the investigation of the effect of imperfections on the physical and structural properties of crystals is well documented \cite{As76, Ki76, Ch00}. The effects of imperfections/defects on hyperuniformity have also been extensively investigated theoretically in the context of perturbed lattices, e.g., see Refs. \cite{We80, Ga04a, Ga04b, Ki18, Kl20} and references therein. Recently, in a seminal study, using various theoretical models, Kim and Torquato \cite{Ki18} demonstrated that while thermal excitation and point defects such as vacancies and interstitials destroy hyperuniformity, uncorrelated random displacements preserve hyperuniformity, but could change the class of hyperuniformity. To quantify the degree of hyperuniformity of real systems, the hyperuniformity index \cite{Ma17b, Ki18, Kl18, Ch18c}
\begin{equation}
\label{eq_4}
H\equiv S(k\rightarrow0)/S(k_{max}) 
\end{equation}
based on the structure factor $S(k)$ is often employed, where $k_{max}$ is the position of the largest peak in the Fourier space. We note that for most practical purposes effective hyperuniform systems with $H\leq10^{-4}$ \cite{Ma17b, Ki18, Kl18, Ch18c} behave essentially the same as perfect hyperuniform systems. However, systematic study of how the introduction of imperfections affect hyperuniformity of real experimental systems, in particular atomic-scale low-dimensional materials, is still lacking.

Very recently, disordered hyperuniformity (DHU) has been observed in the context of various atomic-scale two-dimensional (2D) materials \cite{Zh13, Ge19, Zh20, Ch20}. For example, DHU is found to arise in patterns of electrons emerging from a quantum jamming transition of correlated many-electron state in a quasi-two-dimensional dichalcogenides, which leads to enhanced electronic transport \cite{Ge19}. It is also discovered that DHU distribution of localized electrons in 2D amorphous silica results in an insulator-to-metal transition in the material, which is in contrast to the conventional wisdom that disorder generally diminishes electronic transport \cite{Zh20}. Moreover, Chen and coworkers \cite{Ch20} have rigorously demonstrated that the introduction of Stone-Wales (SW) topological defects \cite{St86} into honeycomb network structures preserves hyperuniformity of these systems to a large extent, and the resulting amorphous structural models capture the salient features of graphene-like 2D materials at low temperatures.

While SW defects are prevalent in graphene-like 2D materials, other types of defects such as chalcogen vacancies, metal vacancies and trefoil defects are dominant in monolayer transition metal dichalcogenides (TMDCs) such as MoS$_2$ and WSe$_{2}$ \cite{Li15, Le20}. For example, Lin and coworkers \cite{Li15} have elaborated how the local structures evolve as various types of defects are introduced into MoS$_2$. There are also a few preliminary experimental studies \cite{Le19c, Le20b, Ko13} examining the evolution of structures as various types of defects are introduced into samples of TMDCs.  

In this work, we conduct a comprehensive characterization of the evolution of global structures as defects are gradually introduced into real experimental samples of TMDCs, in the context of hyperuniformity.  Specifically, we employ deep-learning algorithms to extract the atomic positions in a sequence of image frames obtained from  ADF-STEM, as the scanning electron probe continuously introduces defects into a monolayer crystalline 2D transition metal dichalcogenide alloy, Te-doped 2H-WSe$_2$. We then employ a multi-step procedure to identify and refine the chemical-bonding informed coordination networks in this evolving system. Subsequently, we employ a variety of theoretical and quantitative tools from soft-matter physics, in particular pair statistics and bond-orientation statistics to quantify the evolution of global structures. We find that the systems are nearly hyperuniform at low defect concentrations, and the (approximate) hyperuniformity is completely destroyed even when there is still a significant portion of crystalline sites (specifically, less than $20 \%$ defects). No intermediate hexatic phase are found to exist as the defects are gradually introduced into the systems, which is distinctly different from the 2D melting process as temperature increases. Moreover, we generalize an analytical formula to describe the structure factor $S(k)$ of crystal with correlated displacements and vacancies. Using this analytical formula and Monte Carlo simulations, we demonstrate that the experimental samples in the early frames can be essentially viewed as perturbed crystals with small correlated displacements and double chalcogen vacancies. Moreover, our results indicate the level of degradation of hyperuniformity that one should expect due to the finite experimental measurement precision in real STEM experiments. In addition, we note that our analysis procedures can be readily adapted to characterize the structures of other ordered and disordered two-dimensional materials.

\begin{figure}[h!]
\begin{center}
$\begin{array}{c}\\
\includegraphics[width=0.40\textwidth]{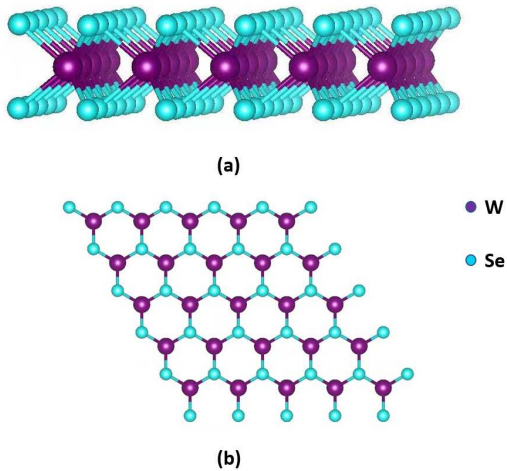} 
\end{array}$
\end{center}
\caption{(Color online) A schematic (a) of the 2H-WSe$_2$ monolayer and its projection (b) when seen from above. Note that the Te-doped 2H-WSe$_2$ monolayer (at low Te concentration can be viewed as effectively the 2H-WSe$_2$ monolayer), when projected from above, is mapped into a perfect honeycomb lattice (ignoring the local strains introduced by the Te substitutions, which are small compared to our measurement precision), with each ``particle'' in the projected plane possessing three bonds. Moreover, half of the honeycomb lattice sites are occupied by the W atoms, and each of the other half sites is occupied by 2 overlaying Se/Te atoms. Each pair of W sites are separated by a Se/Te site, and vice versa.} \label{fig_1}
\end{figure}

The rest of the paper is organized as follows:
in Sec. II, we describe the methods that we employ to extract atomic coordinates and determine chemical bonds 
between atoms from the obtained STEM images, as well as the statistical descriptors that we use to characterize these structures. 
In Sec. III, we employ various statistical descriptors to characterize the evolving global structures of the experimental samples. In Sec. IV, we present a minimalist structural model of the real experimental samples in the early frames, and use analytical formula and Monte Carlo simulations to validate it. In Sec. V, we provide concluding remarks.

\begin{figure}[h!]
\begin{center}
$\begin{array}{c}\\
\includegraphics[width=0.50\textwidth]{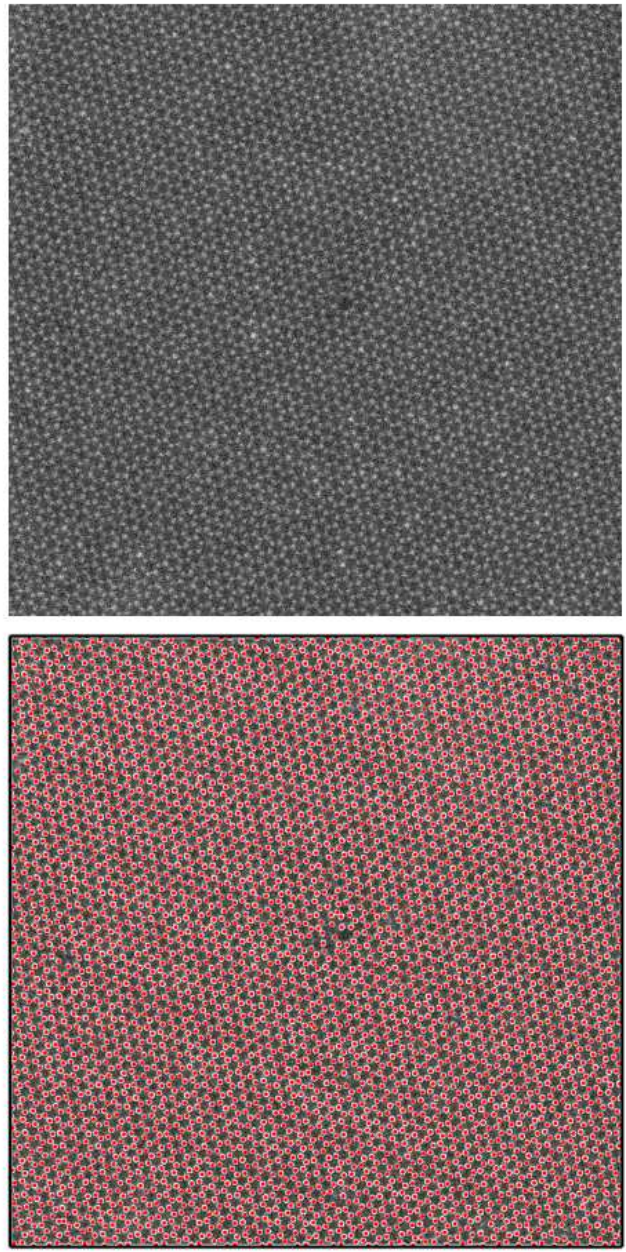} 
\end{array}$
\end{center}
\caption{(Color online) A schematic illustrating the mapping from a raw image (top) of a defected monolayer crystalline transition metal dichalcogenide obtained using the ADF-STEM technique to the extracted atomic positions (bottom).} \label{fig_2}
\end{figure}

\begin{figure*}[t]
\begin{center}
$\begin{array}{c}\\
\includegraphics[width=0.995\textwidth]{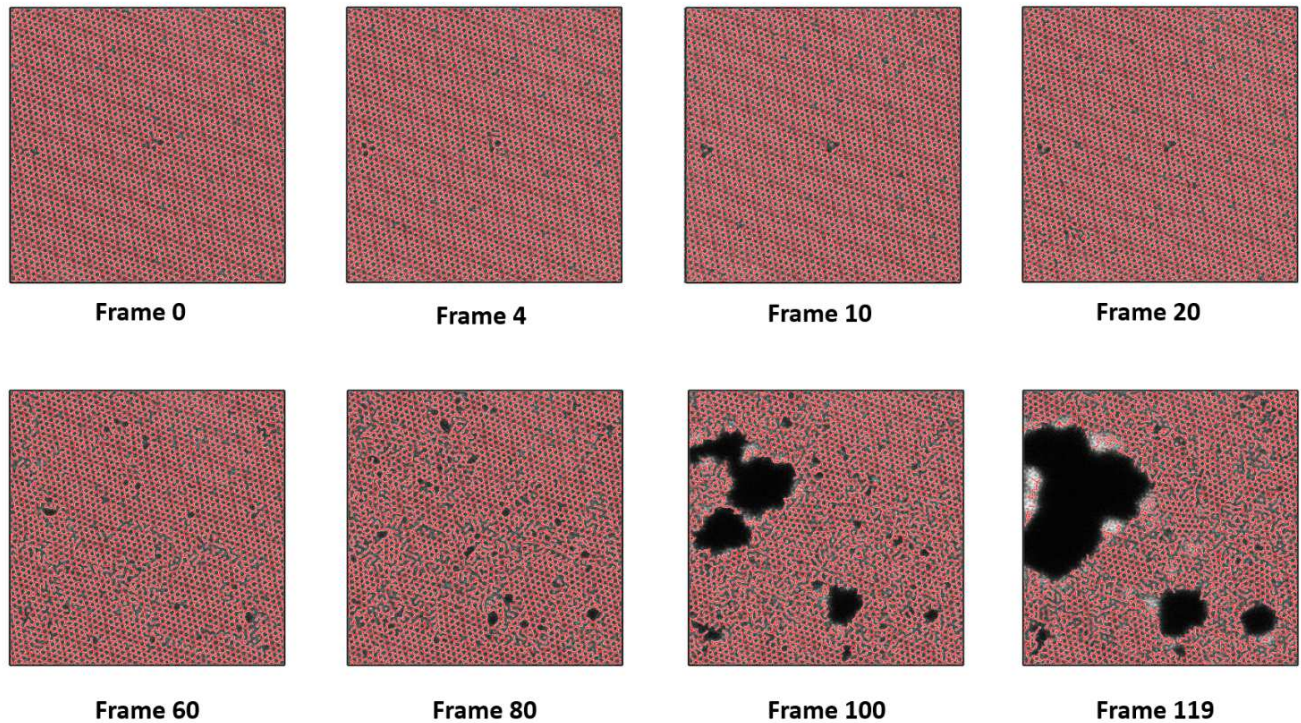} 
\end{array}$
\end{center}
\caption{(Color online) Extracted atomic coordinates and determined chemical-bonding informed coordination networks overlaid with raw images of different frames obtained using the ADF-STEM technique.} \label{fig_3}
\end{figure*}

\section{Methods}
\subsection{Extraction of Atomic Coordinates and Identification of Chemical-Bonding Informed Coordination Networks}
We acquire aberration-corrected scanning transmission electron microscopy images of Te-doped WSe$_2$. In ADF-STEM, an angstrom-scale electron beam is scanned across the sample, and scattered electrons are collected as a function of the position of the electron beam. The material studied is WSe$_{2-2x}$Te$_{2x}$ where $x = 0.06$. The synthesis and TEM sample preparation methods were previously described in Ref. \cite{Le20}. Because the Te fraction is small and the local lattice distortion induced by Te substitutions (2-4 pm) is below the measurement precision of these frames ($\sim$15 pm), we do not expect the impact of the Te to be significant (a schematic of the 2H-WSe$_2$ structure is shown in Fig. \ref{fig_1}), and the sample can be treated as if it were primarily WSe$_2$. During imaging, beam-induced defects, primarily vacancies, voids, and local stripes of phase transformations \cite{Li14}, gradually modify the underlying lattice. In this text, we analyze a 120-frame atomic-resolution movie to capture the generation and evolution of beam-induced defects.
The readers are referred to the supplementary information of Ref. \cite{Le20} for more details including sample fabrication, acquisition parameters, and the movie itself. Next, we extract 2D-projected atomic coordinates from the movie using a deep learning package (AtomSegNet \cite{Li20}). The extracted atomic coordinates are further processed to remove possible artifacts from the imaging and the deep learning treatment, particularly for the atomic coordinates that are too close to each other. 
For example, we merge together any group of atoms that are within 73.3 pm from each other by averaging over their coordinates. Since the average shortest projected bond length is around 190 pm, pair of atoms within the 73.3 pm threshold are considered non-physical and thus merged together. The mapping from a raw image frame to the extracted atomic positions is schematically shown in Fig. \ref{fig_2}.

To construct chemical-bonding informed coordination networks from the final extracted atomic positions, we compute the distance between each pair of atoms, and use a three-step procedure to identify and refine the chemical bonds. We first assign a bond to those pairs with a pair distance smaller than the cutoff 241.89 pm, which is found to identify the bonds in the crystalline region relatively well. Next, we add bonds around the defects by allowing a larger cutoff of 322.52 pm for any particle with one or zero bonds identified in the first step. Finally, we limit the maximum number of bonds that any particle can possess to three and remove the extra bonds by sorting bonds according to the bond length and only keeping the shortest three bonds for each particle. Visual examination indicates that overall our procedure generates reasonably accurate coordination networks that faithfully represent the actual chemical bonding network. 

\subsection{Statistical descriptors}
We consider a variety of statistical descriptors that are sensitive in picking up structural information across length scales. A basic quantity of a point configuration is the aforementioned pair correlation function $g_2(r)$, which is proportional to the probability density function of finding two centers separated by distance $r$ \cite{To02a}. In practice, $g_2(r)$ is computed via the relation
\begin{equation}
\label{eq_5}
 g_2(r) = \frac{\langle N(r)\rangle}{\rho 2\pi r\Delta r},
\end{equation}
where $\langle N(r)\rangle$ is the average number of particle centers that fall into the circular ring at distance $r$ from a central particle center (arbitrarily selected and averaged over all particle centers in the system), $2\pi r\Delta r$ is the area of the circular ring, and $\rho$ is the number density of the system \cite{To02a, At13}. The static structure factor $S(k)$ is the Fourier counterpart, and for computational purposes, $S({k})$ is the angular-averaged version of $S({\bf k})$, which can be obtained directly from the particle positions ${\bf r}_j$, i.e.,
\begin{equation}
\label{eq_6}
S({\bf k}) = \frac{1}{N} \left |{\sum_{j=1}^N \exp(i {\bf k} \cdot
{\bf r}_j)}\right |^2 \quad ({\bf k} \neq {\bf 0}),
\end{equation}
where $N$ is the total number of points in the system \cite{Za09, At13, Ch14}. The trivial forward scattering contribution (${\bf k} = 0$) in Eq. \ref{eq_6} is omitted, which makes Eq. \ref{eq_6} completely consistent with the aforementioned definition of $S(k)$ in the ergodic infinite-system limit.

The aforementioned local number variance $\sigma_N^2(R)$ is another quantity that is often used to characterize density fluctuations of many-body systems. To compute $\sigma_N^2(R)$, we randomly place circular observation windows with radius $R$ in the system under the constraint that the windows should fall entirely within the image frame to avoid boundary artifacts \cite{Ji14, Ch16}. Also, the largest radius of the window that one can sample must be much smaller than half of the box length, otherwise density fluctuations are artificially diminished \cite{Dr15}. We count the number of particles $N(R)$ that fall into the observation window, which is a random variable. The variance associated with $N(R)$ is denoted by $\sigma_N^2(R)\equiv\langle N(R)^2\rangle-\langle N(R)\rangle^2$, which measure density fluctuations of particles within a window of radius $R$.

The bond-orientational order metric $Q_6$ and correlation function $C_6(r)$ \cite{Za99, Wi11} are often used to study the melting process. Specifically, the order metric $Q_6$ is defined as 
\begin{equation}
\label{eq_7}
Q_6\equiv |\langle \Psi_6 \rangle|, 
\end{equation}
where 
\begin{equation}
\label{eq_8}
\Psi_6({\bf r}_i) = \frac{1}{n_i}\sum_{j=1}^{n_i}e^{6\theta_{ij}},
\end{equation}
and $\langle\cdots\rangle$ denotes ensemble average, $n_i$ is the number of neighbors of vertex $i$ located at ${\bf r}_i$, and $\theta_{ij}$ is the polar angle associated with the vector from vertex $i$ to $j$-th chemically bonded neighbor of vertex $i$. 

The bond-orientational correlation function $C_6(r)$ is defined as 
\begin{equation}
\label{eq_9}
C_6(r)\equiv\langle \Psi_6({\bf r}_i)\Psi^*_6({\bf r}_j)\rangle\mid r=|{\bf r}_i-{\bf r}_j|,
\end{equation} 
where $\Psi^*_6$ is the complex conjugate of $\Psi_6$. In practice, to compute $C_6(r)$, for each pair of particles located at ${\bf r}_i$ and ${\bf r}_j$, respectively, we compute $\Psi_6({\bf r}_i)\Psi^*_6({\bf r}_j)$, and bin the results according to the distance $r=|{\bf r}_i-{\bf r}_j|$. We note that $Q_6 = 1$ and $C_6(r) = 1$ for a perfect honeycomb network; while for isotropic fluid phase, $Q_6 \approx 0$ and $C_6(r)$ decays with an exponential envelop at large $r$ \cite{Za99, Wi11}. To avoid artifacts caused by the image boundaries, we exclude the vertices that are within certain distance (439.8 pm in this work) from each edge of the bounding box of the image.

\section{Structural characterization of evolving global structure of defected two-dimensional transition metal dichalcogenides}

We apply the aforementioned scheme to extract atomic coordinates from different frames of STEM images, as shown in Fig. \ref{fig_3}. We then construct chemical-bonding informed coordination networks using the aforementioned procedures and the resulting networks are shown in Fig. \ref{fig_3}. We note that crystalline Te-doped 2H-WSe$_2$, an alloyed 2D TMDC monolayer, when projected from above onto a 2D plane, is mapped into a perfect honeycomb lattice and thus hyperuniform, with each ``particle'' in the projected plane possessing three bonds. Moreover, half of the honeycomb lattice sites are occupied by the W atoms, and each of the other half sites is occupied by 2 overlaying Se/Te atoms. Visual examination of the configurations and the associated networks in Fig. \ref{fig_3} suggest that the samples in the early frames appear to be primarily affected by double chalcogen vacancies. The AtomSegNet \cite{Li20} identifies any occupied site with significant image intensity as an atom.  As a result, it does not distinguish between metal and chalcogen sites, and it only identifies a site as a ``defect''  if there is no atom in projection or the intensity is significantly weaker than others. For example, although single chalcogen vacancies are also prevalent \cite{Le20} in these Te-doped 2H-WSe$_2$ samples, these vacancies still contain single chalcogen atoms and will not be identified as defects in the \textit{projected} structures. Also, the atomic coordinates in the crystalline region deviate slightly from the perfect honeycomb crystal, which might be due to various experimental factors including 1) detector noise, 2) uncertainties introduced by the deep-learning algorithm, and 3) instrument instabilities (sample drift, scanning errors, mechanical vibrations). These factors limit the measurement precision and accuracy to $\sim15$ pm. It is noteworthy that various types of defects such as vacancies and interstitials, and correlated displacements may destroy or degrade hyperuniformity of the structures, as studied theoretically. In the later frames, large voids begin to form in the samples. The Te-doped WSe$_2$ sample gradually evolves from a nearly perfect crystal to a highly defective crystal due to the damage introduced by the electron beam used for imaging. The high energy (80keV) electrons induce knock-on damage and radiolysis in the sample, producing vacancies, voids, and local lattice reconstruction. The ability to image and generate atomic-scale defects allows us to systematically study how hyperuniformity evolves with defect concentration. In addition, we note that when analyzing any real experimental samples from images, one is almost inevitably limited by the finite precision for the determination of atomic coordinates, which effectively adds random uncorrelated displacements to the particle positions. However, as previously proven in a theoretical study, random uncorrelated displacements preserve hyperuniformity, so finite measurement precision should not affect our ability of determining the particular hyperuniformity property of a given experimental system.

\subsection{Pair statistics}
\begin{figure*}[t]
\begin{center}
$\begin{array}{c}\\
\includegraphics[width=0.995\textwidth]{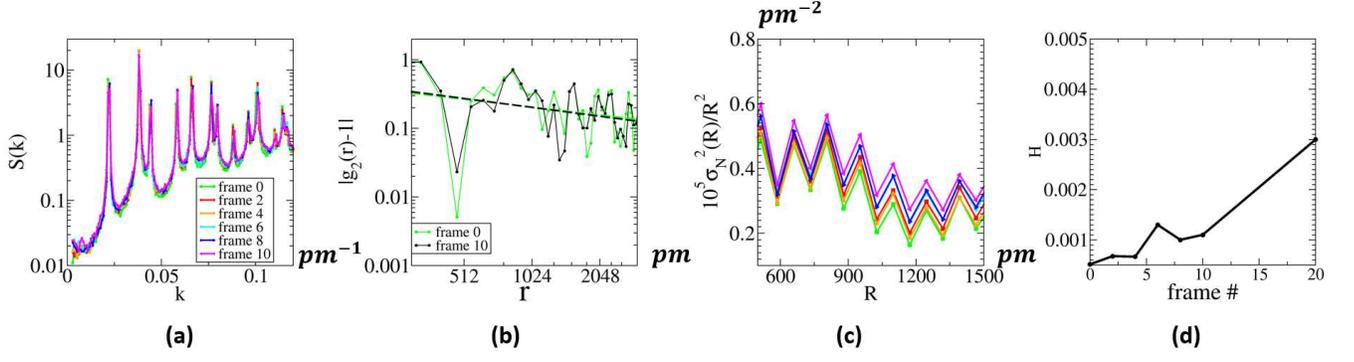} 
\end{array}$
\end{center}
\caption{(Color online) Pair statistics associated with different frames of ADF-STEM images in the nearly hyperuniform regime. (a) Structure factor $S(k)$. (b) Log-log plot of $|g_2(r)-1|$ with a linear fitting (dashed line). (c) Normalized local number variance $\sigma_N^2(R)/R^2$. The legends in (c) are the same as those in (a). (d) Hyperuniformity index $H$.} \label{fig_4}
\end{figure*}

\begin{figure*}[t]
\begin{center}
$\begin{array}{c}\\
\includegraphics[width=0.995\textwidth]{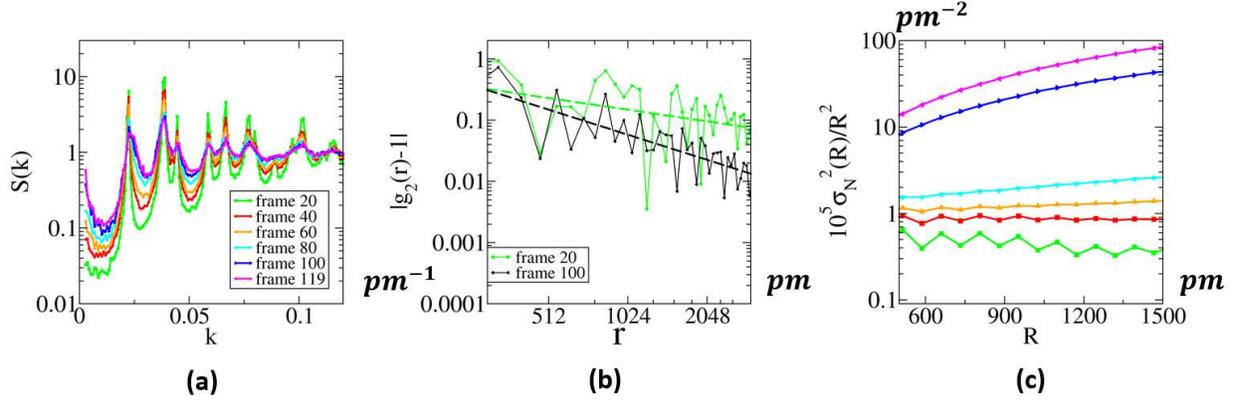} 
\end{array}$
\end{center}
\caption{(Color online) Pair statistics associated with different frames of ADF-STEM images in the nonhyperuniform and antihyperuniform regime. (a) Structure factor $S(k)$. (b) Log-log plot of $|g_2(r)-1|$ with a linear fitting (dashed line). (c) Normalized local number variance $\sigma_N^2(R)/R^2$. The legends in (c) are the same as those in (a).} \label{fig_5}
\end{figure*}

While previously it has been rigorously demonstrated that random introduction of even a tiny fraction of vacancies into a crystal destroys \textit{perfect} hyperuniformity [i.e., $S(k\rightarrow0)$ is strictly zero] of the crystal, the quantification of the degree of approximate hyperuniformity of real experimental samples when vacancies and other types of imperfections are jointly affecting the structures remains an important problem to explore, which we address in the following sections. It is noteworthy that there appear to be correlations between damage events, which are initially low (when the vacancy concentration is low) and then increase (e.g., when large voids begin to form). To characterize the density fluctuations of the experimental samples across different length scales, we compute various pair statistics of these samples, and the results are shown in Figs. \ref{fig_4} and \ref{fig_5}. In particular, as the defects are gradually introduced into the system from frame 0 to frame 20, the magnitudes of the Bragg peaks in $S(k)$ decrease, which indicates the degradation of the crystalline order of the system. Moreover, the structure factor $S(k)$ of the structures in frames 0-10 appear to decrease slightly or plateau to relatively small values as $k$ approaches 0, indicating suppressed large-scale fluctuations. On the other hand, $S(k)$ of frame 20 in Fig. \ref{fig_5} appears to converge to value appreciably larger than zero as $k$ decreases at small $k$, showing the destruction of hyperuniformity at this point. There are significant wiggles in $S(k)$ at large $k$ as well in all of these structures, suggesting short-scale structures in these materials. Also, the scaling exponent $\beta$ in $|h(r)|=|g_2(r)-1| \sim1/r^{\beta}$ increases from frame 0 to frame 10, i.e., the total correlation function $h(r)$ decays faster as $r$ increases as defects are introduced into the system, which is consistent with the increasing disorder and loss of large-scale structural correlation in the system. In addition, the normalized local number variance $\sigma_N^2(R)/R^2$ of the structures in frames 0-10 decreases slightly as $R$ increases at large $R$, indicating (approximate) hyperuniformity of the structures, while $\sigma_N^2(R)$ scales as $R^2$ for frame 20, indicating the loss of hyperuniformity at this point. To quantify the degree of hyperuniformtiy, we employ the hyperuniformity index $H$ for the series of structures in different frames by extrapolating $S(k)$ to $k=0$ with a linear fitting of $S(k)$ within $k\in[0.0025 pm^{-1}, 0.0147 pm^{-1}]$. We find that $H\lessapprox 10^{-3}$ for the first 10 frames and $H>10^{-3}$ for frame 20, suggesting that the structures in frames 0-10 are nearly hyperuniform and the (approximate) hyperuniformity is essentially destroyed for the structure in frame 20. We note that the hyperuniformity index $H$ is primarily suited for characterizing hyperuniform systems or those that are not too away from being hyperuniform, and thus we only compute $H$ for frames 0-20.

For structures in frames 20-40, $S(k)$ plateaus at small $k$ and converges to finite values as $k$ approaches 0, and $\sigma_N^2(R)$ scales as the volume of the observation window (i.e., $R^2$) as shown in Fig. \ref{fig_5}, indicating that the corresponding systems enter the nonhyperuniform regime. In this regard, these structures are similar to typical liquids and glasses \cite{Ha86, To18a}. As more defects are introduced and large voids begin to appear beyond frame 60, $S(k)$ of the corresponding structures start to diverge as $k$ goes to zero, and $\sigma_N^2(R)$ grows faster than the window volume (i.e., $\sigma_N^2(R)/R^2$ is an increasing function of $R$), and such structures can be regarded as antihyperuniform with hyperfluctuations \cite{To18a}, since they are the antithesis of a hyperuniform system. Note that antihyperuniform systems are often observed at thermal critical points, all of which have fractal structures \cite{To18a}. To our best knowledge, this is the first time that antihyperuniformity is observed in real disordered atomic-scale 2D materials. In addition, the total correlation function $h(r)$ decays faster as $r$ increases as more defects are introduced into the system beyond frame 20, which is consistent with the increasing disorder.

\subsection{Bond-orientational statistics}
\begin{figure*}[t]
\begin{center}
$\begin{array}{c}\\
\includegraphics[width=0.98\textwidth]{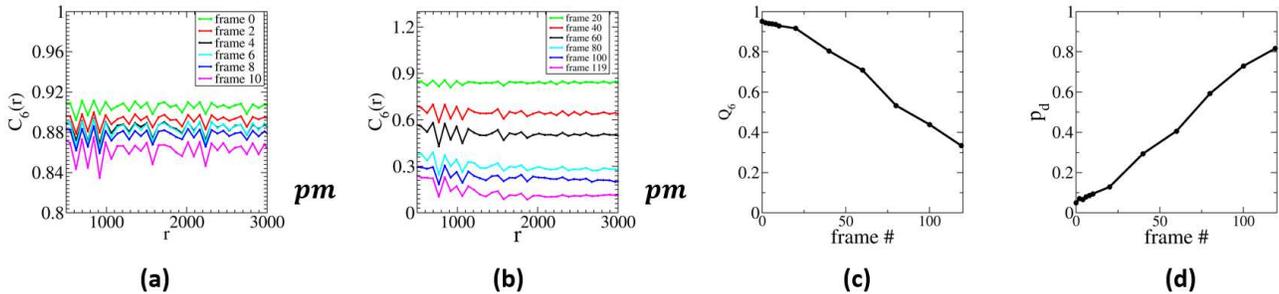} 
\end{array}$
\end{center}
\caption{(Color online) Bond-orientational statistics associated with different frames of ADF-STEM images. (a) Bond-orientational correlation function $C_6(r)$ of the frames in the nearly hyperuniform regime. (b) Bond-orientational correlation function $C_6(r)$ of the frames in the nonhyperuniform and antihyperuniform regime. (c) Bond-orientational order parameter $Q_6$. (d) Defect fraction $p_d$.} \label{fig_6}
\end{figure*}

The process of introducing defects into the honeycomb lattice can also be viewed as the ``melting'' of hexagonal 2D materials to some extent. It is widely known that 2D melting of colloidal systems is a two-step crystalline-hexatic-liquid phase transition, and the bond-orientational correlation function changes from oscillating around certain constant to power-law scaling, and then exponential scaling at large length scale as temperature increases \cite{Ko73, Za99, Xu08, Wi11}. To investigate the melting behavior of our experimental system, we compute the bond-orientational order metric $Q_6$ and bond-orientational correlation function $C_6(r)$, which have been routinely used to study 2D melting \cite{Za99, Wi11}. The results of $C_6(r)$ and $Q_6$ are shown in Fig. \ref{fig_6}. It can be clearly seen that $C_6(r)$ oscillates around certain constant for all of the investigated structures, i.e., the scaling behavior does not change, although that constant decreases as defects are gradually introduced, suggesting decreased orientational order of the system. Moreover, $Q_6$ also decreases relatively smoothly, which is consistent with the results of $C_6(r)$. We note that $Q_6$ is small but still much larger than 0 even for the structure in frame 119, which means that there is remaining orientatinal order in the system, a reflection of the presence of the remaining crystalline regions in the system. These results indicate that there is no intermediate hexatic emerging in the ``melting'' of 2D TMDC monolayer, which is consistent with the fact that there is no mechanism in the system leading to the formation of grain boundary and the loss of long-range orientational order. This behavior is different from the 2D melting of colloidal systems and similar to the observation in structural models of graphene-like materials where disorder is introduced through the SW topological defects \cite{Ch20}. 

To quantify the ``perfectness'' of the honeycomb lattice during the damaging process, we compute the defect concentration $p_d$ defined as:
\begin{equation}
    p_d = 1 - N_c/N_0,
\end{equation}
where $N_c$ is the number of crystalline sites in a structure (excluding the region within 439.8 pm of the edges), and $N_0$ 
is the number of particles in a reference perfect honeycomb lattice (if it were to occupy the same region) where the side length of a hexagon in the perfect crystal is set as the same as the average bond length of all the crystalline sites in the experimental structure under consideration. Here we consider a particle to reside in a crystalline site if the following conditions are met: 1) this particle has three bonds; 2) the bond length difference of its longest bond and shortest bond is within certain threshold (set as 36.65 pm here); 3) all of its bond angles are in the vicinity of 120 degrees (set as $100\sim140$ degrees here). This geometric definition of defect concentration is distinct from the ``real'' atomic point defect concentration in the lattice (for example the number of Se vacancies, Te substitutions, or antisite defects). Visual examination of identified crystalline sites in different frames indicates that our procedure produces reasonably good results consistent with our definition of crystalline sites. We note that combining the results in Figs. \ref{fig_4}-\ref{fig_6}, we find that the (approximate) hyperuniformity is essentially destroyed when the defect fraction $p_d$ is much smaller than 20$\%$, i.e., when the material still contains a significant amount of crystalline sites, and the structures enter the antihyperuniform regime when $p_d$ exceeds 40$\%$.

\section{Structural model of defected two-dimensional transition metal dichalcogenides}
As mentioned above, the experimental samples in the early frames appear to be primarily affected by double chalcogen vacancies and correlated displacements. To fully understand the effect of imperfections on the structures, we construct a simplistic structural model of the experimental samples that enables us to tune different factors independently and see the outcomes. Here we consider a simple honeycomb
lattice where each particle is connected to its chemically-bonded neighbors by springs of spring constant $K$. Specifically, the energy $E$ of the system is given by
\begin{equation}
\label{eq_10} E=\sum_{i<j,H_{ij}=1}\frac{Ka^2}{2}|\frac{\mathbf{u}_i-\mathbf{u}_j}{a}|^2,
\end{equation}
where $H_{ij}=1$ indicates that vertices $i$ and $j$ are connected by a chemical bond, and $\mathbf{u}_i$ is the vector displacement of vertex $i$ from its corresponding reference honeycomb lattice site. We introduce correlated displacements to the particles in the honeycomb lattice according to a multivariate Gaussian distribution:
\begin{equation}
    p(\mathbf{u}) \sim \exp(-\frac{E}{k_BT}) = \prod_{i<j,H_{ij}=1} \exp(-\frac{|(\mathbf{u}_i-\mathbf{u}_j)/a|^2}{2\sigma^2})
\end{equation}
where $\sigma^2=k_BT/(Ka^2)$ is an effective variance associated with the Gaussian distribution and can be viewed as ``fictitious'' dimensionless temperature. Here $k_B$ is the Boltzmann's factor, $T$ is the ``fictitious'' temperature, and $a$ is the side length 
of the hexagon in the original honeycomb lattice. By varying $T$, one can effectively vary the variance $\sigma^2$. It is noteworthy that uncorrelated stochastic displacements of crystals preserve perfect hyperuniformity, and for this reason we employ correlated displacements in our model to account for the loss of perfect hyperuniformity observed in the relatively defect-free experimental samples in the first few frames. Experimentally,  correlated displacements can result from as scan distortions, sample drift, mechanical vibration, noise, and local phase transitions. However, thermal motion can be excluded from the potential source of these displacements since the time scale associated with thermal motion is much shorter than the STEM image acquisition time, and thus the obtained atomic positions are time-averaged and the thermal motion information filtered out. 

Note that introducing correlated displacements in the above fashion is mathematically equivalent to generating equilibrium structures at finite positive temperature, which would allow us to utilize standard Monte Carlo simulations to obtain perturbed honeycomb lattices with correlated displacements according to Gaussian distribution with different variance $\sigma^2$. Specifically, at each trial move, each vertex is allowed to randomly move within a prescribed maximal distance from its old position in each dimension and the trial move is accepted with the probability
\begin{equation}
\label{eq_11} p_{acc}(old\rightarrow new) = \textnormal{min}\{1, \textnormal{exp}(-\frac{E_{new}-E_{old}}{k_BT})\},
\end{equation}
where $E_{old}$ and $E_{new}$ are the energies of the system before and after the trial move as defined in Eq. \ref{eq_11}. To fully equilibrate the system, we perform $1000N$ trial moves, where $N$ is the number of particles/vertices in the system. 

\begin{figure}[h!]
\begin{center}
$\begin{array}{c}\\
\includegraphics[width=0.40\textwidth]{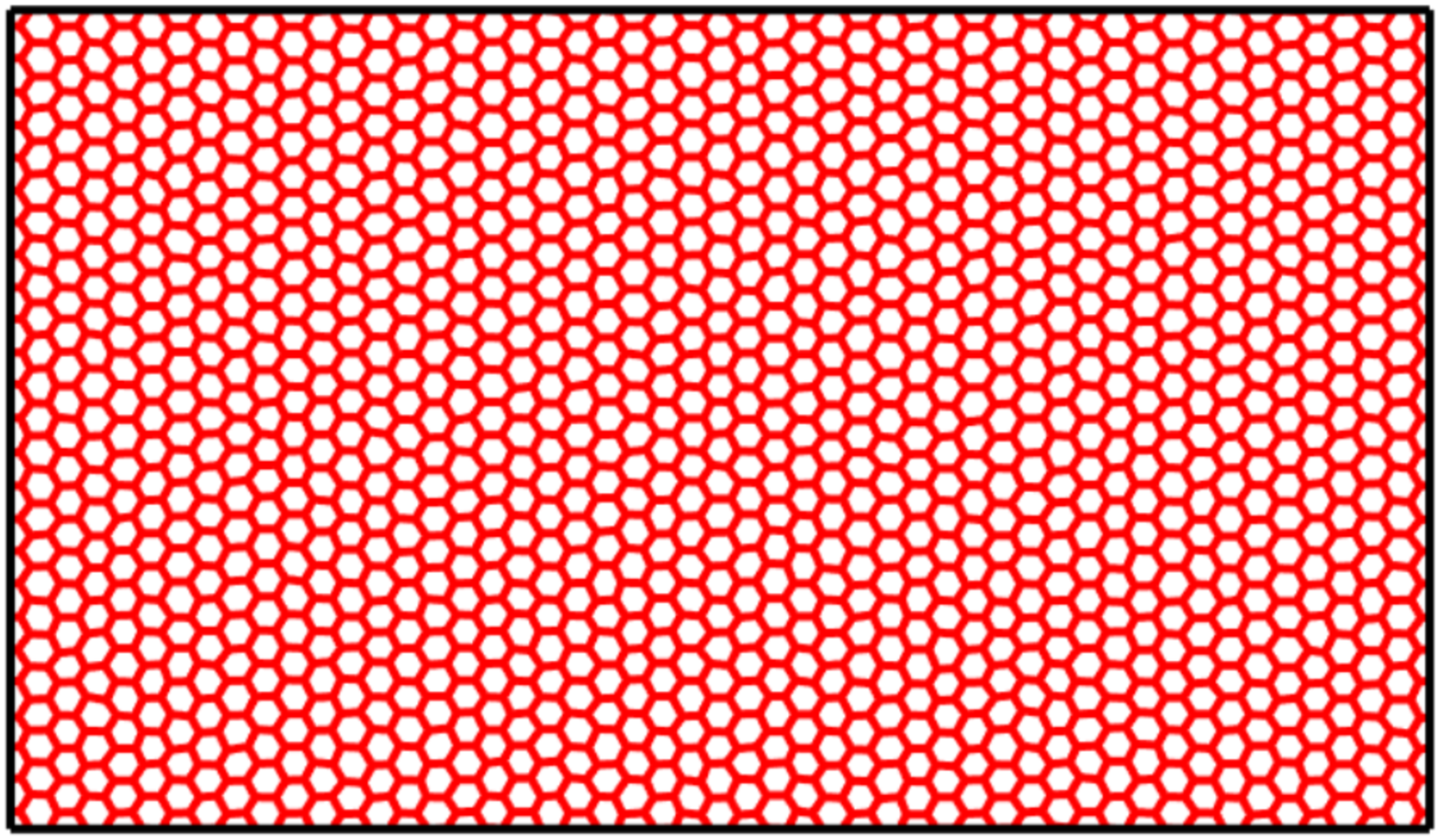} \\
\mbox{\bf (a)} \\
\includegraphics[width=0.40\textwidth]{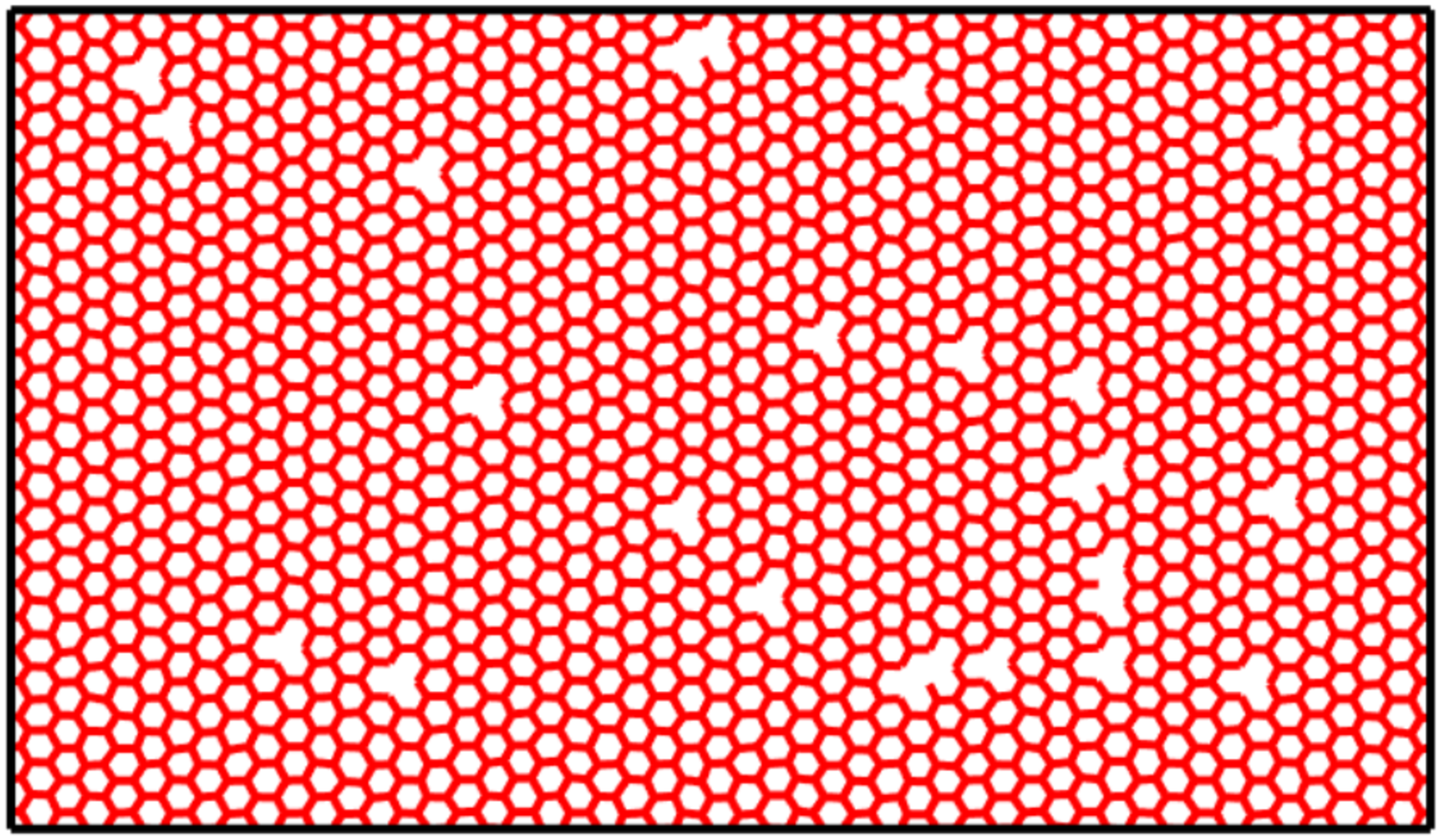} \\
\mbox{\bf (b)} \\
\includegraphics[width=0.40\textwidth]{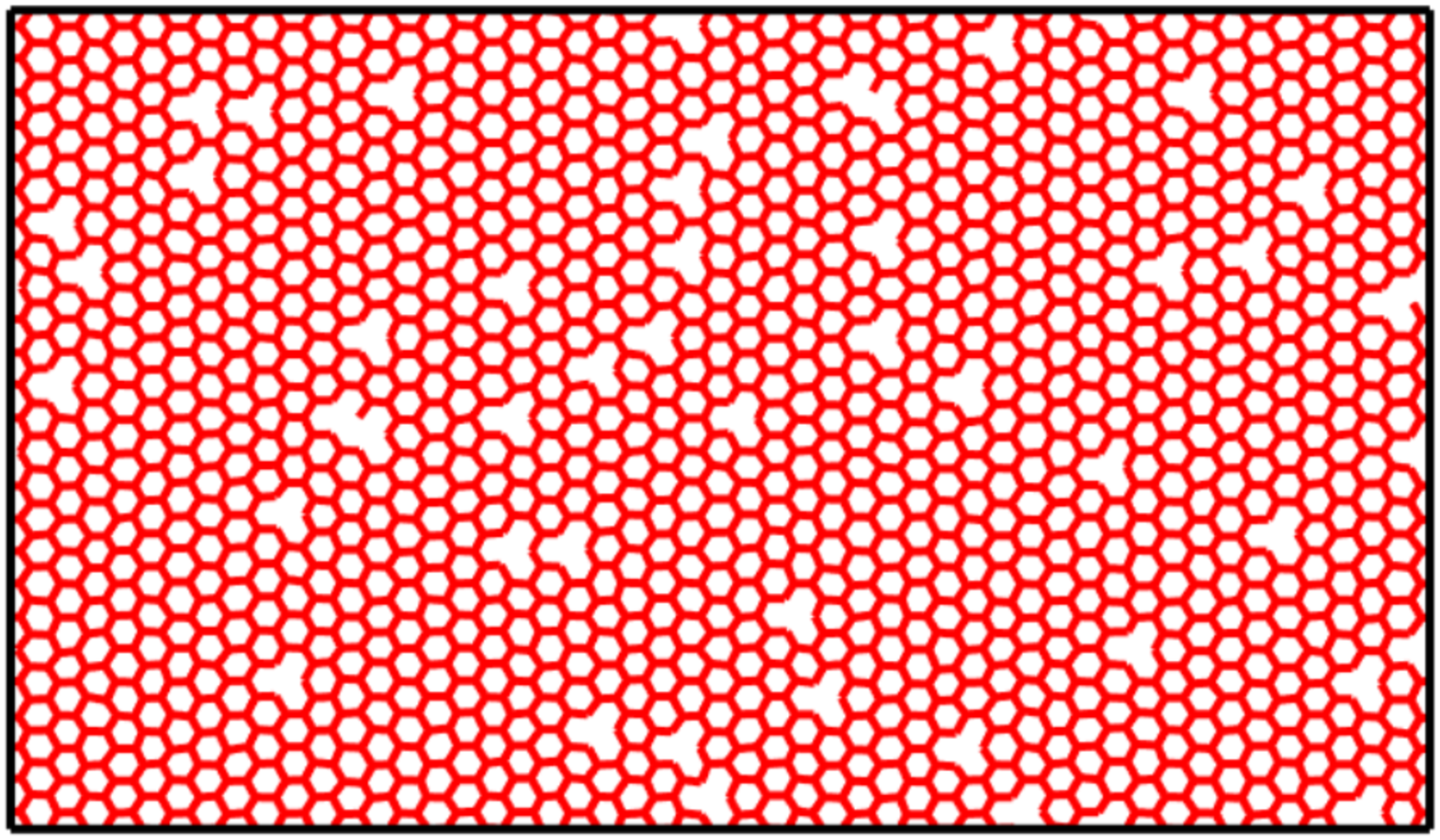} \\
\mbox{\bf (c)} 
\end{array}$
\end{center}
\caption{(Color online) Representative configurations of honeycomb crystal with correlated Gaussian displacements with variance $\sigma^2=0.005$ at different vacancy concentrations $p$. (a) $p=0$. (b) $p=0.02$. (c) $p=0.04$.} \label{fig_7}
\end{figure}

Using this scheme, we generate configurations of displaced honeycomb crystals with $N=2,500$ particles at different $\sigma^2$. Moreover, to introduce chalcogen vacancies into the displaced crystal, we randomly remove $pN/2$ particles at chalcogen sites according to prescribed vacancy concentration $p$, where $p$ is defined as the ratio of double chalcogen vacancies over the total number of chalcogen sites in the lattice. Representative configurations at $\sigma^2=0.005$ with varying $p$ are shown in Fig. \ref{fig_7}. We note that chalcogen sites comprise half of the total sites in the honeycomb lattice, and every pair of chalcogen sites is separated by a metal site. We then compute $S(k)$ of these stochastically displaced configurations with or without defects. To better compare the results with the experimental systems, we normalize the distance in our simulated systems so that the characteristic length scale $d_0=2\pi/k_0$ is the same for the initial experimental frame and the vacancy-free displaced crystal at a given temperature, where $k_0$ is the position of the first peak.

\begin{figure}[h!]
\begin{center}
$\begin{array}{c}\\
\includegraphics[width=0.40\textwidth]{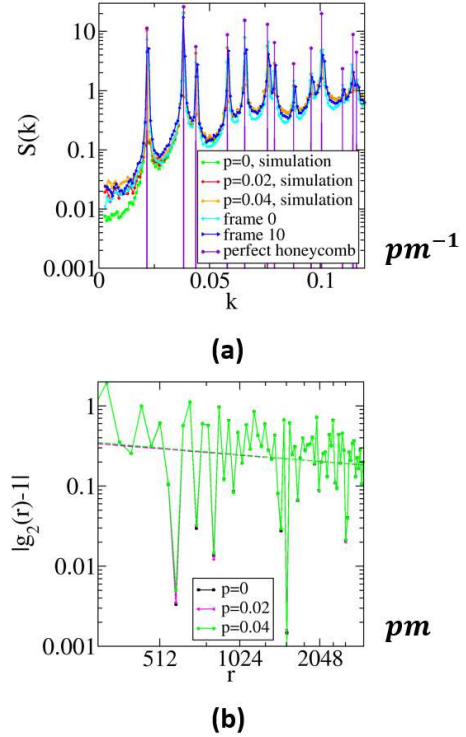} 
\end{array}$
\end{center}
\caption{(Color online) Pair statistics of simulated honeycomb lattice with correlated Gaussian displacements with variance $\sigma^2=0.005$ at different vacancy concentrations. (a) Structure factor $S(k)$ of the simulated frames and representative reference experimental frames, and reference perfect honeycomb lattice.} (b) Log-log plot of $|g_2(r)-1|$ with a linear fitting (dashed line). \label{fig_8}
\end{figure}

We find that $S(k)$ of vacancy-free displaced honeycomb crystal at $\sigma^2=0.005$ shown in Fig. \ref{fig_8}(a) captures the salient features of the experimental sample in the initial frame, which contains relatively few vacancies. Also note here that correlated displacements already largely degrades hyperuniformity even in the absence of vacancies, similar to the effect 
of thermal motions \cite{Hu47, De73} . Moreover, as more chalcogen vacancies are introduced into the simulated system, $S(k)$ at small $k$ further increases as $p$ increases, as shown in Fig. \ref{fig_8}(a), indicating the loss of hyperuniformity. In addition, the scaling exponent $\beta$ in $|h(r)|=|g_2(r)-1| \sim1/r^{\beta}$ slightly increases as $p$ increases, i.e., the total correlation function $h(r)$ decays faster as $r$ increases as vacancies are introduced into the system, as shown in Fig. \ref{fig_8}(b). These trends are consistent with those observed in experimental systems, and demonstrate that the series of experimental samples studied in the early frames can be essentially viewed as honeycomb crystals with small correlated displacements and varying concentration of double chalcogen vacancies. In Fig. \ref{fig_8}(a) we also show $S(k)$ of the reference perfect honeycomb crystal with $N=2,500$ particles in Fig. \ref{fig_7}(a), and $S(k)$ is strictly zero in the intervals between the sharp Bragg peaks.

\begin{figure}[h!]
\begin{center}
$\begin{array}{c}\\
\includegraphics[width=0.40\textwidth]{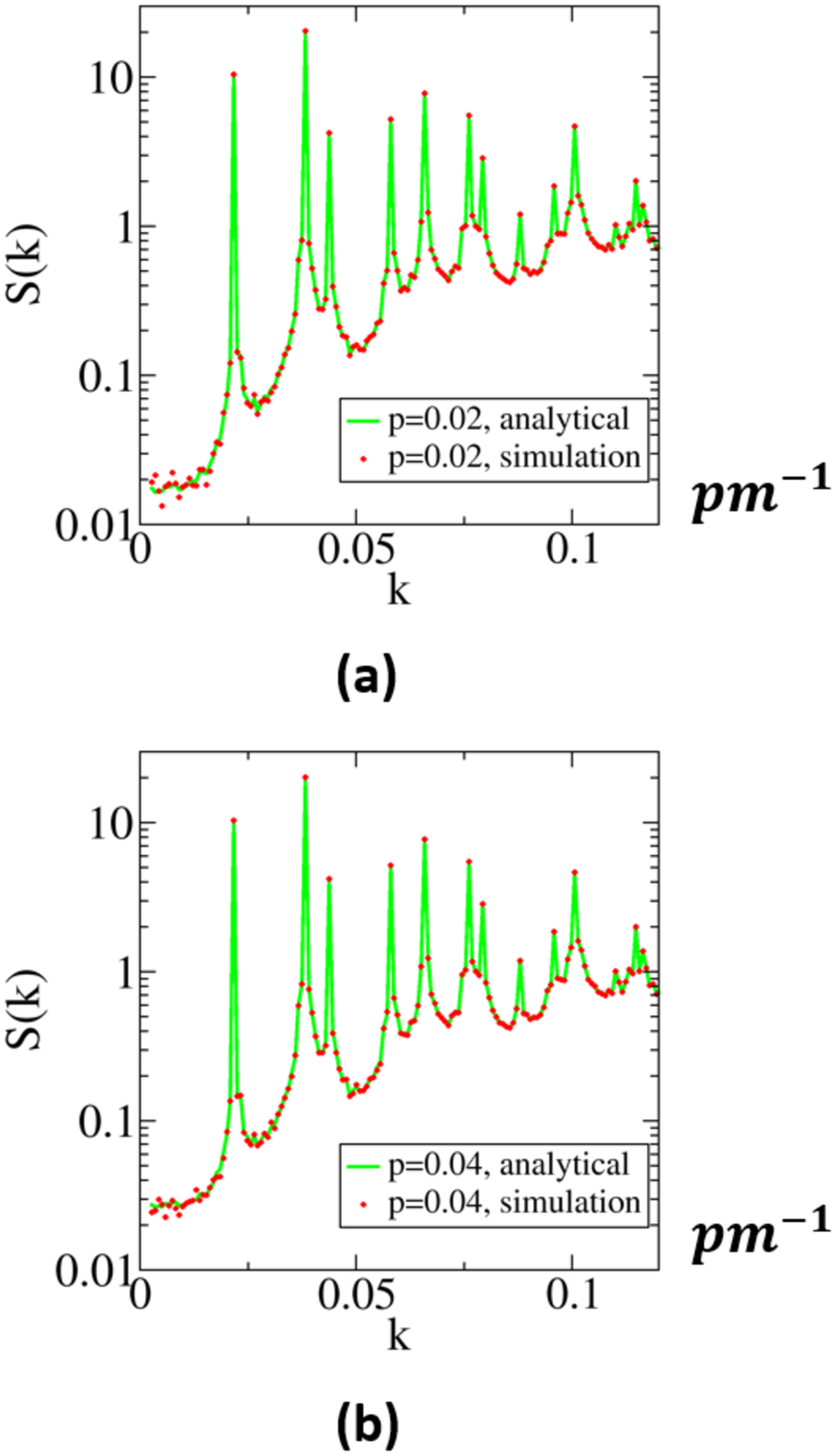} 
\end{array}$
\end{center}
\caption{(Color online) Structure factor $S(k)$ of honeycomb crystal with correlated Gaussian displacements with variance $\sigma^2=0.005$ at different vacancy concentrations obtained from analytical expression (solid line) and Monte Carlo simulations (dots). (a) $p=0.02$. (b) $p=0.04$.} \label{fig_9}
\end{figure}

To further validate our simulation results, we generalize a previously known analytical formula for $S(k)$ of defective point process in any space dimension with spatially uncorrelated point vacancies \cite{Ki18} to displaced honeycomb crystals with double chalcogen vacancies. We note that this formula was first known for defective crystals \cite{Pe76, De73}. Specifically, following similar procedures in Ref. \cite{Ki18} , we derive the generalized formula for $S(k)$ as
\begin{equation}
\begin{split}
  S(k)&=1+(1-p')[S_0(k)-1],\\
    &=1+(1-\frac{p}{2})[S_0(k)-1],\\
    &=\frac{p}{2} + (1-\frac{p}{2})S_0(k),
\end{split}
\end{equation}
where $S_0(k)$ is the structure factor of the vacancy-free displaced crystal, and $p'=p/2$ is the effective vacancy concentration in the system, i.e., the ratio of chalcogen vacancies over the total number of lattice sites. Since in our model vacancies can only be introduced randomly at chalcogen sites, strictly speaking the vacancies are not completely spatially uncorrelated, but we assume that this spatial correlation can be neglected. We compute $S(k)$ of displaced honeycomb crystal with $p$ concentration of double chalcogen vacancies using this analytical formula by plugging in the simulated $S_0(k)$ of vacancy-free displaced crystals. We find that the analytical results match the directly simulated $S(k)$ of defected displaced crystals well, as shown in Fig. \ref{fig_9}, which demonstrates that our assumption is indeed valid. 

\begin{figure}[h!]
\begin{center}
$\begin{array}{c}\\
\includegraphics[width=0.40\textwidth]{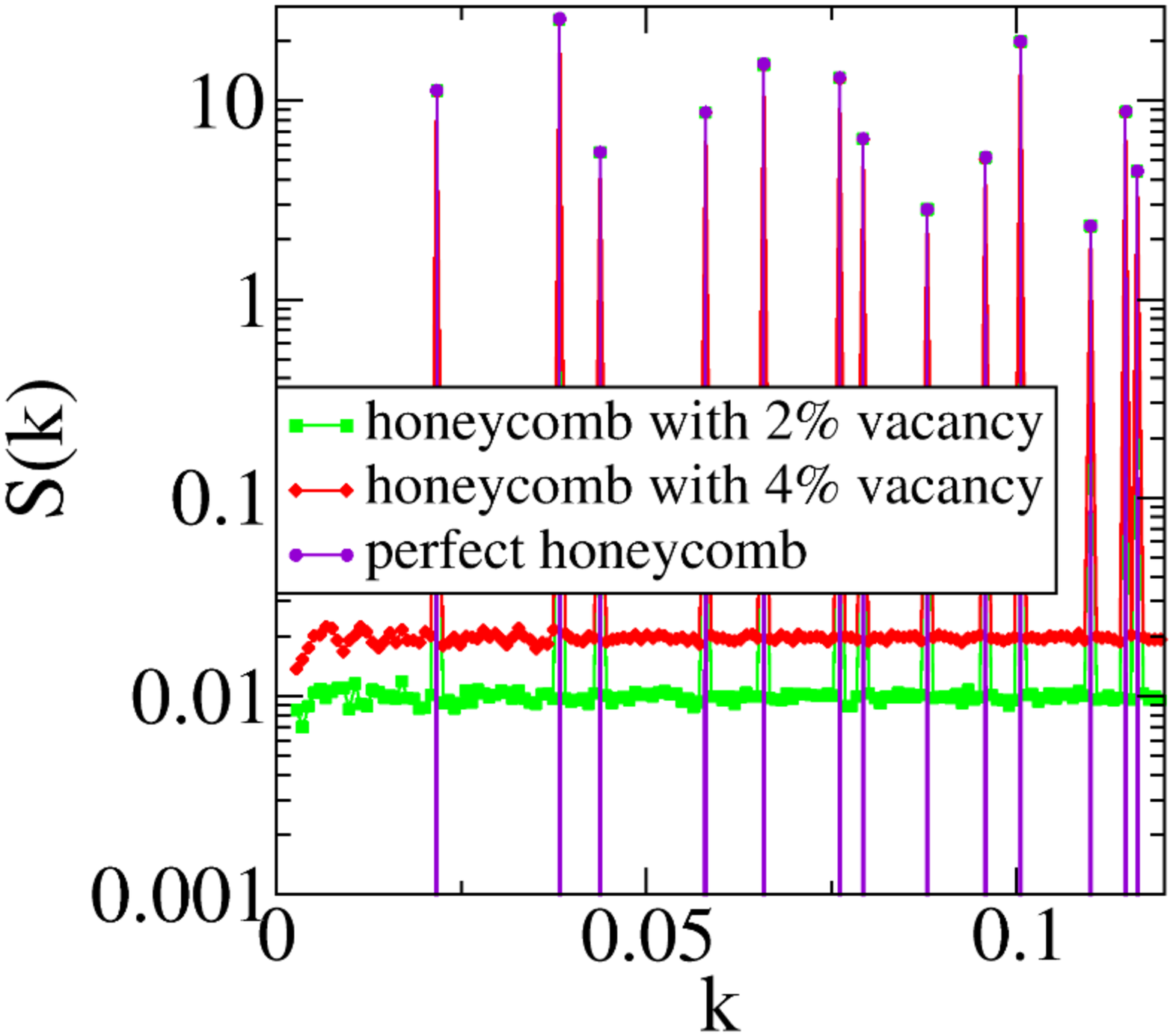} 
\end{array}$
\end{center}
\caption{(Color online) Structure factor $S(k)$ of honeycomb crystals at different double chalcogen vacancy concentrations without correlated displacements.} \label{fig_10}
\end{figure}

Because certain correlated displacements of atomic positions may be due to various experimental measurement errors rather than intrinsic features of the experimental samples, we also use computer simulations to investigate the large-scale structural features of honeycomb crystals with double chalcogen vacancies and no correlated displacements. Specifically, we randomly remove particles at chalcogen sites and compute the structure factor $S(k)$ of the defected honeycomb crystals at different double chalcogen vacancy concentrations without correlated displacements, as shown in Fig. \ref{fig_10}. It is noteworthy that $S(k)$ is relatively flat away from the Bragg peaks associated with the honeycomb crystals, and the values of $S(k)$ in these intervals between Bragg peaks are roughly proportional to the effective vacancy concentration, which is consistent with previous theoretical predictions \cite{Ki18}.

\section{Conclusion and Discussion}
In this work, we investigated structural features of timelapse STEM images of 2D TMDCs across length scales as an electron probe was used to gradually introduce various types of defects into the 2D materials. In particular, we quantified density fluctuations and the associated hyperuniformity/antihyperuniformity property of these defected 2D materials. We find that in the early frames the chemical bonding-informed coordination network is mainly influenced by double chalcogen vacancies, and at very low defect concentrations the corresponding materials are nearly hyperuniform, as manifested in various pair statistics and quantified by the hyperuniformity index $H$. However, as additional defects are introduced, the (approximate) hyperuniformity of the materials is completely destroyed, when there is significant amount of crystalline regions in the system. No intermediate hexatic phase emerged, which is different from the 2D melting process for colloidal systems. In later frames large voids begin to form in the samples, leading to the rise of antihyperuniformity of the structures. By constructing a minimalist structural model for the samples in the early frames, we were able to demonstrate that the experimental samples can be essentially viewed as perturbed honeycomb crystals with small correlated displacements and double chalcogen vacancies. Moreover, the correlated displacements alone, which is the result of various uncontrollable experimental noises and perturbations that one should usually expect when acquiring STEM images of atomic-scale 2D materials, largely degrade hyperuniformity of the system, and low concentration of vacancies, when coupled with correlated displacements, basically destroy (approximate) hyperuniformity. 

We note that here we primarily studied the effect of double chalcogen vacancies and correlated displacements on the density fluctuations of defected TMDCs. However, if other more complex types of defects such as trefoil defects can be introduced into experimental samples of TMDCs in a controllable manner, in principle we can employ similar procedures to investigate the effect of those defects. Also, it is noteworthy that while chalcogen vacancies are dominant in the projected structures of certain samples of TMDCs such as those in the early frames of the current work, SW defects are prevalent in graphene-like 2D materials. In the future it would be interesting to look at how SW defects, when coupled with other factors, affect structural features and physical properties of real graphene-like materials in experiments. Previously, it has been demonstrated \cite{Ch20} that the introduction of SW defects and local structural relaxation alone preserves hyperuniformity of honeycomb lattice while the disorder of the structure increases, and is accompanied by the emergence of exotic physical properties. It is important to note that while a single experimental realization of amorphous graphene was previously found to be hyperuniform \cite{Ch20}, even the samples of TMDCs in the very early frames in the current work are found to be only nearly (or approximately) hyperuniform. More generally, to our best knowledge we for the first time introduce a variety of theoretical machinery from soft-matter physics to study the structural evolution of experimental samples of atomic-scale 2D materials, and these techniques can be readily adapted and applied to other ordered and disordered 2D materials. The structural studies of 2D materials in this paper and related works will strengthen our fundamental understanding of the physics underlying these materials, and serve as the basis for future functional materials design.  

\begin{acknowledgments}
We are grateful for Dr. JaeUk Kim and Dr. Runze Tang for very helpful discussion. H.Z. thank the start-up
funds from Arizona State University (ASU). Y. J. thanks ASU for support and Peking University for hospitality during his sabbatical leave. C.-H. L. and P. Y. H. acknowledge funding support from the U.S. Department of Energy, Office of Basic Energy Sciences, Division of Materials Sciences and Engineering under award $\#$ DE-SC0020190, for electron microscopy data acquisition and processing. W. Z. acknowledges funding support from the Office of Naval Research (ONR) under grant $\#$ NAVY N00014-17-1-2973.
\end{acknowledgments}

\end{document}